\documentclass{ws-procs9x6adap}
\usepackage[T1]{fontenc}
\usepackage[latin1]{inputenc}
\usepackage{graphics}
\usepackage{eepic}
\usepackage{epsfig}

\usepackage{amssymb}

\begin{document}
\title{Cold deconfined matter EOS through an\\ HTL quasi-particle
model}
\author{Paul Romatschke}
\address{Institut f\"ur Theoretische Physik, Technische Universit\"at Wien,
Austria}
\maketitle

\abstracts{Using quasi-particle models, lattice data can be mapped to
finite chemical potential. By comparing a simple and an HTL
quasi-particle model, we derive the general
trend that a full inclusion of the plasmon effect will give.}

\section{Introduction}

The equation of state (EOS) of 
deconfined matter was determined perturbatively to order
$\alpha^{5/2}_{s}$ in 
QCD\cite{FreedmanMcL:1978,ZhaiKas:1995};
however, due to the large coupling and 
strong renormalization scheme dependencies, perturbative methods are plagued by
bad convergence properties that make them fail in the region of physical 
interest; clearly, non-perturbative approaches are needed.

For zero quark chemical potential $\mu=0$ lattice QCD calculations are 
seemingly up to the task of determining the EOS for a quark-gluon plasma
at finite temperature, and although 
recently\cite{FodorKatSza:2002,ForcrandPhi:2002,Allton:2002zi}
there has been some progress also for non-vanishing $\mu$, the determination
of the EOS for cold dense matter, which is of importance in astrophysical
situations\cite{PeshierKaemSof:1999,PeshierKaemSof:2002,FragaPisSchaf:2001,%
AndersenStr:2002},
is currently not possible by means of
lattice calculations.

As a remedy for the situation, Peshier 
\emph{et al.}\cite{PeshierKaemSof:1999} suggested a method which
can be used to map the available lattice data for $\mu=0$ to finite $\mu$
and small temperatures by describing the interacting plasma as a system
of massive quasi-particles (QPs); section 2 gives a short review of
this technique. While in their simple QP model
the (thermal) masses of the quarks and gluons are approximated
by the asymptotic limit of the hard-thermal-loop (HTL) self-energies, 
we make use of the full, momentum-dependent HTL self-energies in our model 
(which will be presented in section 3). This\cite{BlaizotIanReb:1999,%
BlaizotIanReb:2000} 
amounts to including more of the physically important plasmon-effect
than in the simple QP model, so we expect our result to show the
general trend of the correction that a full
next-to-leading order (NLO) calculation 
of the self-energies (which is work in progress)
would add to the result from Peshier \emph{et al}.
In section 4 we compare the results obtained from the two models
and present our conclusions as well as an outlook in section 5. 

\section{Mapping lattice data to finite chemical potential using QP models}

Assuming a plasma of gluons and $N_{f}$ light quarks
in thermodynamic equilibrium can be described as an ideal gas of 
massive quasi-particles with residual interaction $B$, the pressure 
of the system is given by\cite{PeshierKaemSof:1999}
\begin{equation}
P(T,\mu)=\sum_{i=g,q} p_{i}(T,\mu_{i}(\mu),m_{i}^2)-B(m_{g},m_{q}),
\label{Pressureansatz}
\end{equation}
where the sum runs over gluons (g), quarks and anti-quarks (q) with 
respective chemical potential $0,\pm\mu$; $p_{i}$ are
the model dependent QP pressures and $m_{i}$ are the QP masses, which
are functions of the effective strong coupling $G^2(T,\mu)$.

Using the stationarity of the thermodynamic potential 
under variation of the self-energies and Maxwell's 
relations\cite{PeshierKaemSof:1999}, one obtains
a partial differential equation for $G^2$,
\begin{equation}
a_{T} \frac{\partial G^2}{\partial T} + a_{\mu} \frac{\partial G^2}{\partial %
\mu} = b,
\label{Andrefloweq}
\end{equation}
where $a_{T}$, $a_{\mu}$ and $b$ are coefficients that are given by integrals 
depending on $T,\mu$ and $G^2$.
Given a valid boundary condition, a solution
for $G^2(T,\mu)$ is found by solving the above flow equation by 
the method of characteristics;
once $G^2$ is thus known in the $T,\mu$ plane, the 
QP pressure is fixed completely.
The residual interaction $B$ is then given by the integral
\begin{equation}
B=\int \sum_{i} \frac{\partial p_{i}}{\partial m_{i}^2}\left(\frac{\partial %
m_{i}^2}{\partial \mu} d\mu + \frac{\partial m_{i}^2}{\partial T} dT\right)%
+B_{0},
\label{Bdet}
\end{equation}
where $B_{0}$ is an integration constant that has to be fixed by lattice
data (usually by requiring $P(T_{c},\mu=0)=P_{lattice}(T_{c})$).
Motivated by the fact that at $\mu=0$ and $T\gg T_{c}$ the
coupling should behave as predicted by the perturbative QCD beta-function,
ref. \cite{PeshierKaemSof:1999} used the ansatz
\begin{equation}
G^2(T,0)=\frac{48 \pi^2}{(11N_{c}-2N_{f})\ln{\frac{T+T_{s}}{T_c}\lambda}}.
\label{G2ansatz}
\end{equation}
The parameters $\lambda$ and $T_{s}$ are determined by fitting the entropy 
of the model (which is independent of $B$)
to available lattice data at $\mu=0$. Using (\ref{G2ansatz}) as boundary
condition for (\ref{Andrefloweq}), the above procedure allows one to 
map the lattice EOS from $\mu=0$ to the whole $T,\mu$-plane.

\section{HTL QP model}

Doing a perturbative expansion of the QP contribution to 
the pressure at $\mu=0$ for the simple QP model\cite{PeshierKaemSof:1999}
and comparing to the known result\cite{ZhaiKas:1995}
one finds that while the Stefan-Boltzmann and leading-order interaction
terms are correctly reproduced, 
only $1/(4\sqrt{2})$ of the NLO term (the plasmon effect) 
is included\cite{BlaizotIanReb:2000}.

Whereas in a simple QP model $p_{g}$ and $p_{q}$ in (\ref{Pressureansatz})
are just the free pressure of massive (scalar) bosons and fermions, 
respectively, the 
HTL-resummed entropy\cite{BlaizotIanReb:1999,BlaizotIanReb:2000} leads to
\begin{eqnarray}
p_g & = & - d_{g} \int \frac{d^3 k}{(2\pi)^3} \int _{0}^{\infty}%
 \frac{d\omega}{2\pi} n(\omega)%
\left[2 \rm{Im} \ln{\left(-\omega^2+k^2+\hat{\Pi}_{T}\right)}%
-2\rm{Im} \hat{\Pi}_{T} \rm{Re} \hat{D}_{T} \right. \nonumber \\
& & \left. + \rm{Im} \ln{\left(k^2+\hat{\Pi}_{L}\right)} +%
\rm{Im} \hat{\Pi}_{L} \rm{Re} \hat{D}_{L}\right] \nonumber \\
p_q & = & - d_{q} \int \frac{d^3 k}{(2\pi)^3} \int_{0}^{\infty} %
\frac{d \omega}{2 \pi} \left(f_{+}(\omega)+f_{-}(\omega)\right)%
\left[\rm{Im}\ln{\left(k-\omega+\hat{\Sigma}_{+}\right)} \right. \nonumber \\
& & \left.-\rm{Im}\hat{\Sigma}_{+}\rm{Re}\hat{\Delta}_{+}%
+ \rm{Im}\ln{\left(k+\omega+\hat{\Sigma}_{-}\right)}+%
\rm{Im}\hat{\Sigma}_{-}%
\rm{Re}\hat{\Delta}_{-}\right],
\label{None}
\end{eqnarray}
where $d_{g}=2(N_{c}^2-1)$, $d_{q}=2 N_{c} N_{f}$ for gluons and 
quarks/anti-quarks, respectively; $n(\omega)$ and $f_{\pm}(\omega)$
are the bosonic and fermionic distribution functions and 
$\hat{D}_{T,L}$, $\hat{\Delta}_{\pm}$ 
are the HTL propagators with $\hat{\Pi}_{T,L}$ and $\hat{\Sigma}_{\pm}$
the corresponding self-energies. This increases the included plasmon
effect to $1/4$ of the known value, the remainder coming from NLO
corrections to $\hat{\Pi}$ and $\hat{\Sigma}$.

\vspace*{-0.3cm}
\section{Comparisons}

\begin{figure}
\begin{minipage}[t]{.46\linewidth}
\includegraphics[width=\linewidth]{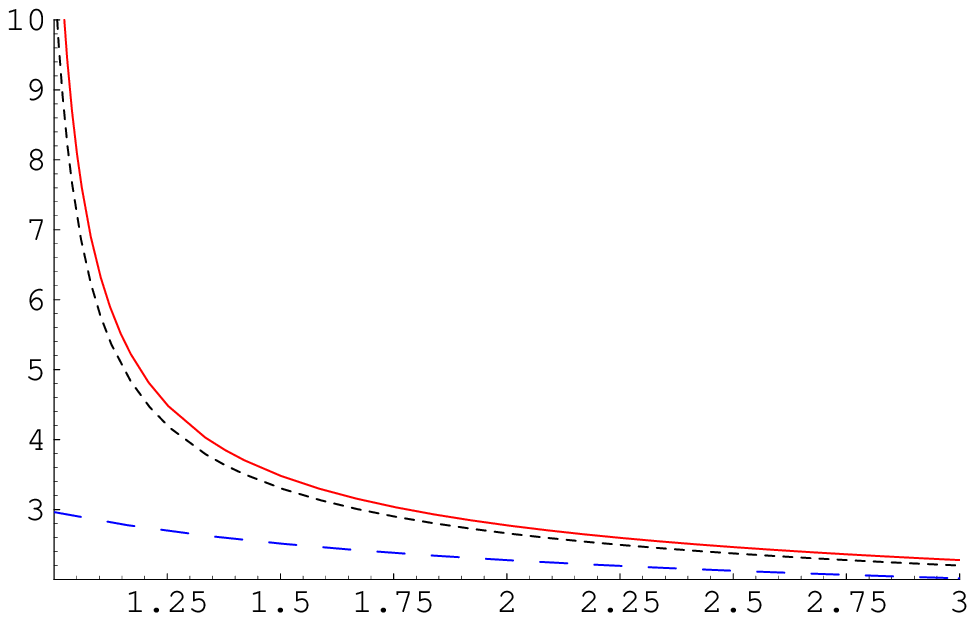}
\setlength{\unitlength}{1cm}
\begin{picture}(6,0)
\put(5,0.1){\makebox(0,0){\footnotesize $T/T_{c}$}}
\put(0.5,4){\makebox(0,0){\footnotesize $G^2$}}
\end{picture}
\caption{$G^2$ at $\mu=0$ for simple
(short-dashed line) and HTL QP model as well as the
2-loop perturbative running coupling\protect\cite{BlaizotIanReb:2000}
in \protect$\overline{\hbox{MS}}$ (long-dashed line).} 
\label{fig:G2mu0comp1}
\end{minipage}\hfill
\begin{minipage}[t]{.46\linewidth}
\includegraphics[width=\linewidth]{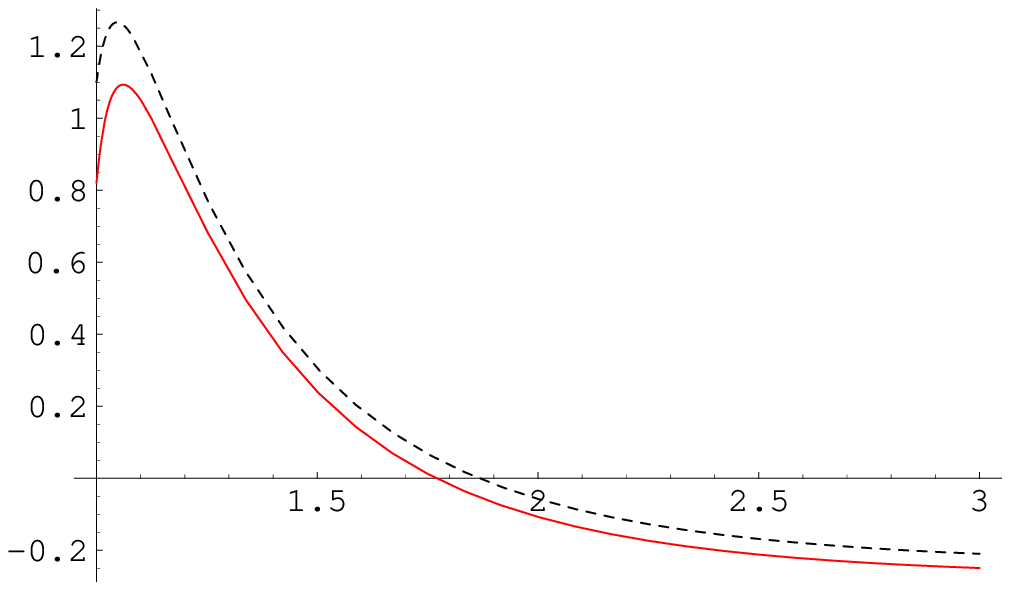}
\setlength{\unitlength}{1cm}
\begin{picture}(5,0)
\put(5,0.1){\makebox(0,0){\footnotesize $T/T_{c}$}}
\put(0.5,4){\makebox(0,0){\footnotesize $B(T,0)/T^4$}}
\end{picture}
\caption{Residual interaction $B/T^4$ at $\mu=0$ for HTL- and simple QP model 
(dashed line).}
\label{fig:Bcomp1}
\end{minipage}\hfill
\end{figure}

To obtain the input parameters $T_{s}$ and $\lambda$ we fitted the entropy
expressions from both models to lattice data\cite{AliKhanEA:2001}
for $N_{f}=2$, normalized and scaled\cite{PeshierKaemSof:2002}. For the 
HTL-model one obtains a slightly higher value of $\lambda^{HTL}=19.4$
than in the simple QP model\cite{PeshierKaemSof:2002}, while $T_{s}$ turns
out to be equal.
Requiring $P(T_{c})=P_{lattice}(T_{c})$ then fixes 
$B_{0}^{HTL}=0.82$; comparisons for $B(T,0)$ and 
the effective coupling $G^2$ are shown in figures \ref{fig:G2mu0comp1} and
\ref{fig:Bcomp1}, respectively.

\begin{figure}
\begin{minipage}[t]{.46\linewidth}
\includegraphics[width=\linewidth]{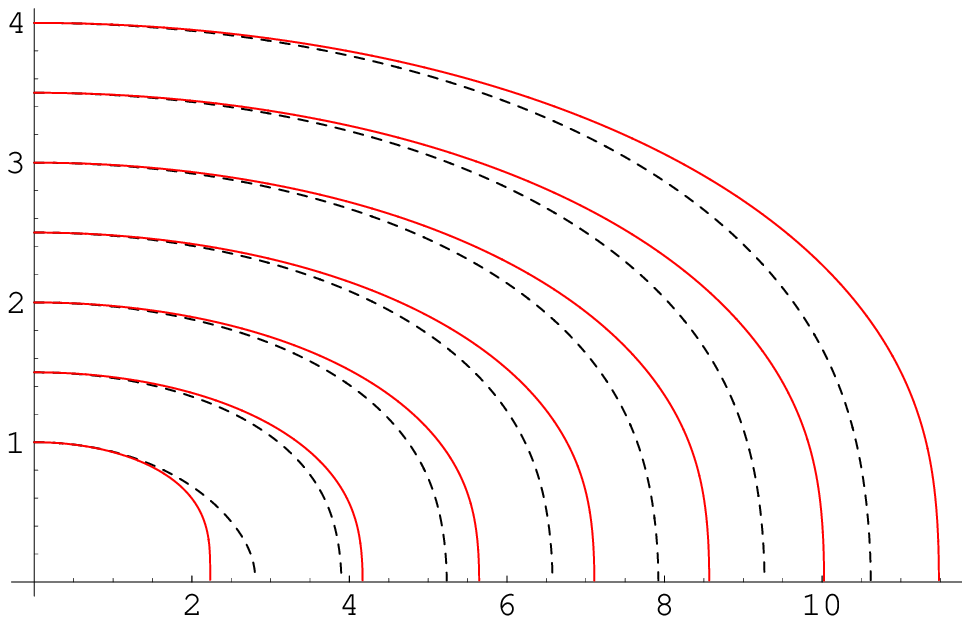}
\setlength{\unitlength}{1cm}
\begin{picture}(6,0)
\put(5,0.1){\makebox(0,0){\footnotesize $\mu/T_{c}$}}
\put(0.5,4){\makebox(0,0){\footnotesize $T/T_{c}$}}
\end{picture}
\caption{Comparison of the shape of the characteristics (dashed lines 
indicate those of the simple QP model).}
\label{fig:charcomp1}
\end{minipage}\hfill
\begin{minipage}[t]{.46\linewidth}
\includegraphics[width=\linewidth]{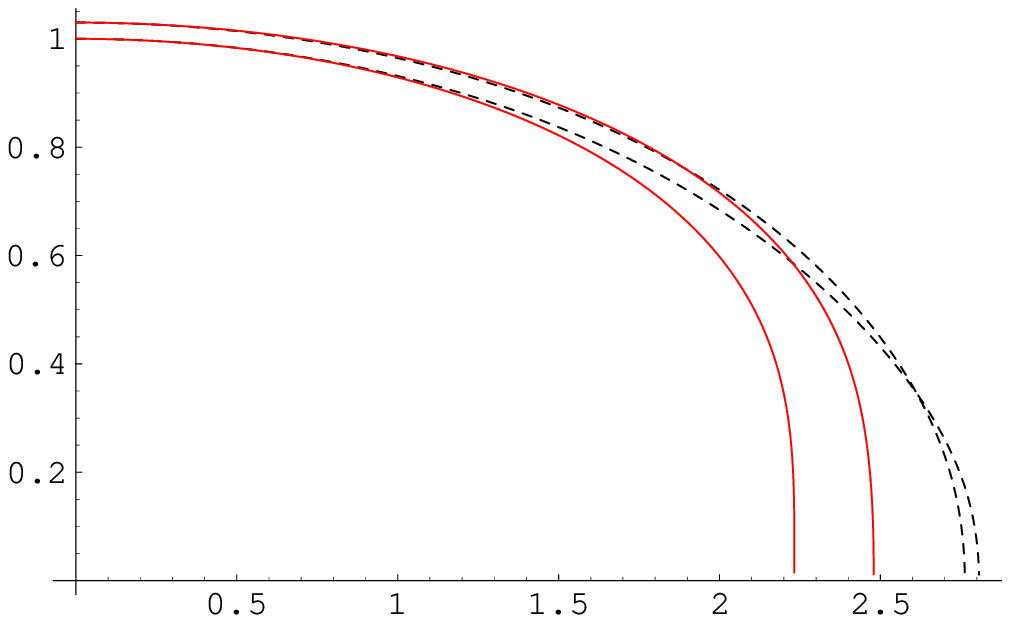}
\setlength{\unitlength}{1cm}
\begin{picture}(6,0)
\put(5,0.1){\makebox(0,0){\footnotesize $\mu/T_{c}$}}
\put(0.5,4){\makebox(0,0){\footnotesize $T/T_{c}$}}
\end{picture}
\caption{Characteristics starting very near $T_{c}$: HTL- and simple
QP model (dashed lines).}
\label{fig:charcomp2}
\end{minipage}\hfill
\vspace*{-0.3cm}
\end{figure}

While the numerical values of the coefficients
$a_{\mu}$ and $a_{T}$ in (\ref{Andrefloweq}) are 
approximately equal in the two models,
the coefficient $b$ turns out to differ noticeably. This difference is 
reflected in the shape of the characteristics (figure
\ref{fig:charcomp1}).
Interestingly enough, 
for characteristics of the simple QP model starting very near
$T_{c}$ intersections occur\cite{PeshierKaemSof:1999}, while for
the HTL model this is not the case (see figure \ref{fig:charcomp2}).

\begin{figure}
\begin{minipage}[t]{.46\linewidth}
\includegraphics[width=\linewidth]{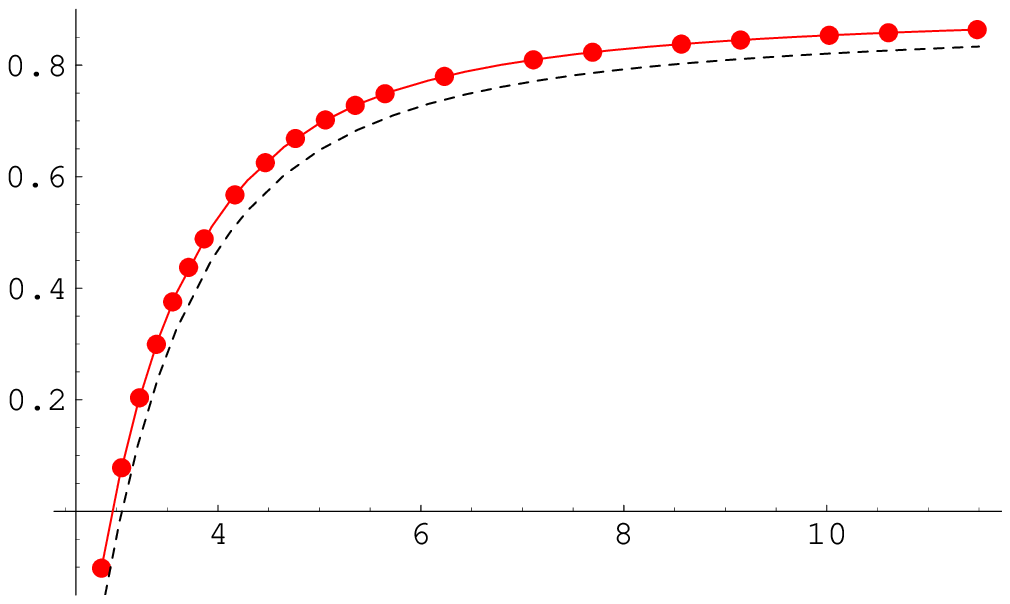}
\setlength{\unitlength}{1cm}
\begin{picture}(6,0)
\put(5,0.5){\makebox(0,0){\footnotesize $\mu/T_{c}$}}
\put(0.5,4){\makebox(0,0){\footnotesize $P/P_{SB}$}}
\end{picture}
\caption{The pressure as a function of $\mu$ at $T=0$ (dashed line 
corresponds to the simple QP model).}
\label{fig:PT0comp1}
\end{minipage} \hfill
\begin{minipage}[t]{.46\linewidth}
\includegraphics[width=\linewidth]{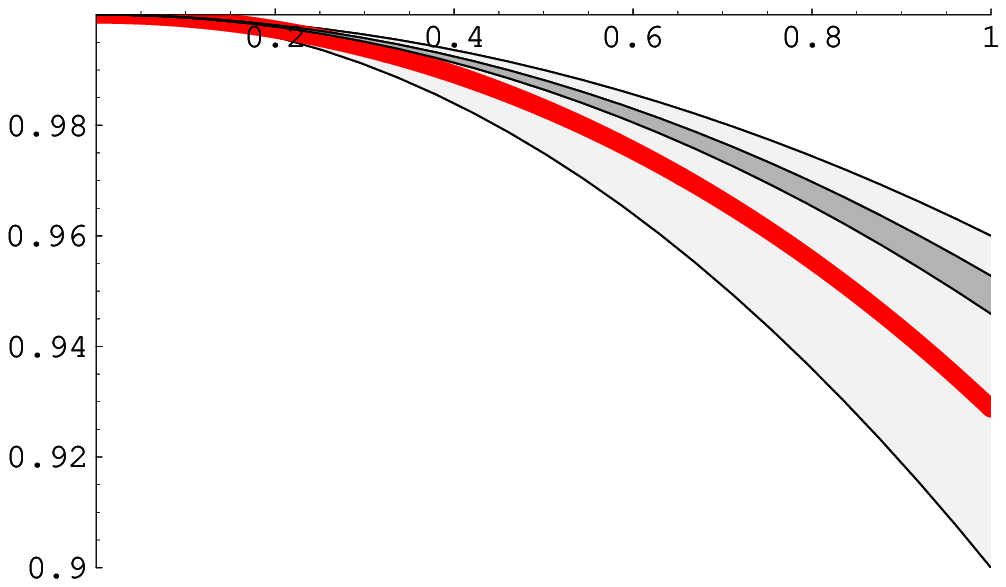}
\setlength{\unitlength}{1cm}
\begin{picture}(6,0)
\put(5,3.9){\makebox(0,0){\footnotesize $\mu/T_{c}$}}
\put(0.5,4){\makebox(0,0){\footnotesize $T/T_{c}$}}
\end{picture}
\caption{Comparison of the QP model characteristics (thick line) to recent lattice
results 
for the phase transition (light\protect\cite{Allton:2002zi} and
dark\protect\cite{ForcrandPhi:2002} gray region).}
\label{fig:CompLat1}
\end{minipage}
\vspace*{-0.3cm}
\end{figure}

Integrating the function $B$ along the characteristics (\ref{Bdet}) 
and adding the QP contributions one obtains
the pressure in the whole $T,\mu$-plane. The result for $T=0$ is shown
in figure \ref{fig:PT0comp1}, where the two curves are simple fits
to the numerical results\cite{HTLtabref}.
Calculating also the
energy density ${\mathcal E}$ for $T=0$ one finds that the
EOS ${\mathcal E}(P)$ for the HTL-model is nearly the same as
for the simple model\cite{PeshierKaemSof:2002}.
Finally, a comparison between the QP model characteristics 
(which are nearly isobars for
small $\mu$) starting at $T=T_{c}\equiv \left.T_{c}\right|_{\mu=0}$ and recent lattice 
data for the phase transition line\cite{ForcrandPhi:2002,Allton:2002zi} 
is shown in figure \ref{fig:CompLat1};
the curvature of the slope for the 
QP model characteristics, $T_{c} dT/d\mu^2\sim-0.07$ coincides with
the central value obtained from one lattice study\cite{Allton:2002zi}.

\section{Conclusions and Outlook}

We have investigated the difference between a simple and an HTL QP model at
finite chemical potential: although the latter incorporates only slightly more
of the plasmon term, there are considerable differences in the characteristics
of the flow equation that eliminate the ambiguities of the solutions for
the simple QP model\cite{PeshierKaemSof:1999}. The differences 
for the thermodynamical quantities are small (even for large $\mu$), which
seems to reflect the fact that these are protected by
the principle of stationarity. However, they
should indicate the trend that an inclusion of the full plasmon term
will give.
It is encouraging that a 
first comparison of the QP models and recent lattice results 
for finite $\mu \lesssim T$ shows some agreement (more extensive studies
are in progress).
\bibliographystyle{wso}
\bibliography{bsample}

\end{document}